\documentclass[prb,showpacs,draft,twocolumn,superscriptaddress]{revtex4}

\usepackage[final]{graphicx}
\usepackage{bm}
\begin{document}

\title{Quantum many-body dynamics in a Lagrangian frame. I. Equations of
motion and conservation laws.}

\author{I.~V.~Tokatly}

\email{ilya.tokatly@physik.uni-erlangen.de}

\affiliation{Lerhrstuhl f\"ur Theoretische Festk\"orperphysik,
  Universit\"at Erlangen-N\"urnberg, Staudtstrasse 7/B2, 91058
  Erlangen, Germany}

\affiliation{Moscow Institute of Electronic Technology,
 Zelenograd, 124498 Russia}
\date{\today}

\begin{abstract}
We formulate equations of motion and conservation laws for a
quantum many-body system in a co-moving Lagrangian reference
frame. It is shown that generalized inertia forces in the co-moving
frame are described by Green's deformation tensor
$g_{\mu\nu}(\bm\xi,t)$ and a skew-symmetric vorticity tensor
$\widetilde{F}_{\mu\nu}(\bm\xi,t)$, where $\bm\xi$ in the Lagrangian
coordinate. Equations of motion are equivalent to those for a
quantum many-body system in a space with time-dependent metric
$g_{\mu\nu}(\bm\xi,t)$ in the presence of an effective magnetic
field $\widetilde{F}_{\mu\nu}(\bm\xi,t)$. To illustrate the
general formalism we apply it to the proof of the harmonic
potential theorem. As another example of application we consider a
fast long wavelength dynamics of a Fermi system in the dynamic
Hartree approximation. In this case the kinetic equation in
the Lagrangian frame can be solved explicitly. This allows us to
formulate the description of a Fermi gas in terms of an effective
nonlinear elasticity theory. We also discuss a relation of our
results to time-dependent density functional theory.
\end{abstract}


\maketitle

\section{Introduction}
Lagrangian and Eulerian formulations of fluid mechanics are known
as two alternative ways to describe dynamics of continuum media
\cite{PhysAc}. The more common Eulerian (or spatial) formulation
considers basic collective variables, such as density $n({\bf
x},t)$ or current ${\bf j}({\bf x},t)$ distributions, as functions
of space-time coordinates ${\bf x}$ and $t$
\cite{PhysAc,LandauVI:e}. This corresponds to the description of a
system from the standard point of view of an observer in a fixed
laboratory reference frame. Central notions of Lagrangian (or
material) description are the trajectories of infinitesimal fluid
elements.  Every small element of a fluid can be uniquely labeled
by its initial position ${\bm \xi}$ that plays a role of
independent, so called Lagrangian, coordinate. Lagrangian
description represents the dynamics of continuum media as it is
seen by a local observer, moving with a flow. In the last decades
the Lagrangian method attracts an increasing attention as a
powerful tool for studying nonlinear dynamics of
compressible media with numerous applications in cosmology, plasma
physics, physics of semiconductors, etc. (for a recent
comprehensive review see Ref.~\onlinecite{Schamel2004}). Recently
we have shown that the Lagrangian coordinate naturally appears in
time-dependent density functional theory (TDDFT), where it plays a
role of a basic variable for a nonadiabatic exchange correlation
potential \cite{TokPPRB2003}. It is also interesting to note a
relation of Lagrangian fluid dynamics to noncommutative geometry and
noncommuting gauge fields \cite{JacPiPol2002}.

Commonly Lagrangian and Eulerian descriptions are considered as
inherent ingredients of the classical continuum mechanics. In fact,
they offer two alternative techniques for solving the equations of
classical hydrodynamics. However, the main idea of Lagrangian
method, which is the description of dynamics using co-moving
coordinates, is clearly much more general and universal. In the
present paper we formulate microscopic equations of many-body
dynamics in the co-moving Lagrangian reference frame. The
transformation to the Lagrangian frame corresponds to an explicit
separation of the convective motion of particles. This is a
natural generalization of the common separation of the center-of
mass motion in homogeneous many-body systems. The
separation of the center-of-mass motion also plays an important role
in the theory of harmonically trapped systems. For the harmonic
inhomogeneity the convective motion can be separated by the
transformation to a global accelerated reference frame, which is a key
step in the proof of the harmonic potential theorem
\cite{Dobson1994,Vignale1995a,Vignale1995b} (HPT). In fact, the proof
of HPT can be viewed as the simplest application of the Lagrangian
description to quantum dynamics. In the case of a general
inhomogeneous flow the separation of convective ``center-of-mass''
motion leads to an appearance of inhomogeneous inertia forces in the
equations for the relative motion. We show that these
forces can be uniquely described by the symmetric deformation
tensor $g_{\mu\nu}(\bm\xi,t)$ and a skew-symmetric vorticity tensor
$\widetilde{F}_{\mu\nu}(\bm\xi,t)$. The deformation tensor enters
equations of many-body dynamics as an effective time-dependent
metric, while the vorticity tensor plays a role of an effective
magnetic field.

A great advantage of the Lagrangian description of many-body dynamics
is that in the co-moving frame both the density of particles and
the current density become the exact integrals of motion. The
current density is zero in every point of Lagrangian
$\bm\xi$-space, while the particles' density distribution preserves
its initial form. These ``conservation laws'' are guaranteed by a
fine local compensation of inertia forces, external forces, and
the force of internal stresses. The above force balance follows
the local momentum conservation law (the exact microscopic
Navier-Stokes equation) after the transformation to the Lagrangian
frame. We explicitly demonstrate that the exact internal stress
force takes a form of a covariant divergence of a symmetric
second-rank stress tensor. As a byproduct of our formalism we
obtain a microscopic representation for the local stress tensor in
a general quantum many-body system.

The concept of quantum stress has been introduced by Schr\"odinger in
1927 \cite{Schrodinger1927}. Over the last two decades there has been
a growing interest in understanding properties of quantum systems,
such as molecules or solids, in terms of the stress density (see, for
example,
Refs.~\onlinecite{BarPar1980,NieMar1985,FilFio2000,RogRap2002} and
references therein). A derivation of a microscopic expression for the
kinetic part of the stress tensor in quantum many-body system causes
no problem. This simple generalization of the
one-particle result has been obtained in the classical paper by Martin and
Schwinger \cite{MartSchw1959}. However, the derivation of the
microscopic form for the interaction related stress tensor turned out
to be not that simple
\cite{MartSchw1959,KadBay,Kugler1967,PufGil1968,Zubarev:e,NieMar1985}.
In this paper we present two alternative derivations of the symmetric
form for the stress density, which has been obtained by Puff and
Gillis in Ref.~\onlinecite{PufGil1968}. In particular, we show that
this form is consistent with the definition of the stress tensor via
the variational derivative of the energy with respect to the metric
tensor.

The structure of the paper is following. In Sec.~II we consider
the standard Eulerian form of the conservation laws in a quantum
many-body system. In this section we also present a compact
derivation of the microscopic expression for the exact stress
tensor. Section~III is devoted to the formulation of quantum
many-body theory in the co-moving Lagrangian frame. In Sec.~IIIA the
key notions of Lagrangian coordinate and of the deformation tensor
are formally defined. The derivation of the equations of motion in an
arbitrary local noninertial reference frame is presented in
Sec.~IIIB. Here we also derive the form of transformed many-body
Hamiltonian and discuss the physical meaning of generalized inertia
forces.  In Sec.~IIIC we derive local
conservation laws, and present a complete formulation of the
many-body problem in the Lagrangian frame. It is shown that this
problem corresponds to the solution of the equations of motion for
the relative motion, supplemented by the local force balance
equation. The force balance equation plays a role of an additional
gauge condition that fixes the reference frame. Section~IV contains
simple examples of application of the general theory. In
Sec.~IVA we interpret the harmonic potential theorem
\cite{Dobson1994} in terms of dynamics in the Lagrangian frame. In
Sec.~IVB we apply the general formalism to the study of
semiclassical collisionless dynamics of a Fermi gas, and shortly
discuss a connection of our approach to TDDFT. It is shown that in
the regime of a fast long wavelength evolution the kinetic
equation in the Lagrangian frame can be solved explicitly. In this
case the behavior of the system is described by an effective
nonlinear continuum mechanics, which, after the transformation to
the laboratory frame, reduces to the generalized collisionless
hydrodynamics of Refs.~\onlinecite{TokPPRB1999,TokPPRB2000}. In
Sec.~V we summarize our results.

\section{Conservation laws in the laboratory reference frame:
  Definition of the stress tensor}

In this paper we consider a system of $N$ interacting particles
in the presence of a time-dependent external potential
$U_{\text{ext}}({\bf x},t)$. The corresponding
Hamiltonian takes the following standard form
\begin{eqnarray}
H &=& \widehat{T}+\widehat{W}+\widehat{U},
\label{1}\\
\widehat{T}&=&
-\int d{\bf x}\psi^{\dag}({\bf x})\frac{\nabla^{2}}{2m}\psi({\bf x}),
\label{2}\\
\widehat{W} &=& \frac{1}{2}\int d{\bf x}d{\bf x'}w(|{\bf x-x'}|)
\psi^{\dag}({\bf x})\psi^{\dag}({\bf x'})\psi({\bf x'})\psi({\bf x}),
\label{3}\\
\widehat{U}&=&
\int d{\bf x}U_{\text{ext}}({\bf x},t)\psi^{\dag}({\bf x})\psi({\bf x}),
\label{4}
\end{eqnarray}
where $w(x)$ is the interaction potential, and $\psi^{\dag}$ and $\psi$
are the field operators, which satisfy proper commutation relations
\begin{equation}
[\psi^{\dag}({\bf x}),\psi({\bf x'})]_{\pm}=\delta({\bf x-x'}).
\label{5}
\end{equation}
The upper (lower) sign in Eq.~(\ref{5}) corresponds to fermions
(bosons), and $[A,B]_{\pm}=AB \pm BA$. Using Hamiltonian of
Eqs.~(\ref{1})--(\ref{4}) we obtain Heisenberg equations of motion
for $\psi$-operators
\begin{eqnarray}\nonumber
i\frac{\partial}{\partial t}\psi({\bf x}) &=&
-\frac{\nabla^{2}}{2m}\psi({\bf x})
+ U_{\text{ext}}\psi({\bf x})\\
&+&\int d{\bf x'}w(|{\bf x-x'}|)
\psi^{\dag}({\bf x'})\psi({\bf x'})\psi({\bf x})
\label{6}
\end{eqnarray}
Equation (\ref{6}) allows to derive equations of motion for any
physical observable as well as for any correlation function. The most
important of these equations are the local conservation laws or
balance equations, which should be satisfied for an arbitrary
evolution of the system. Below we concentrate on conservation laws for
the number of particles and for momentum. These local conservation laws
follow the equations of motion for the density, $n({\bf
x},t)$, and for the current, ${\bf j}({\bf x},t)$ respectively. Computing the
time derivative of the density operator,
\begin{equation}
\widehat{n}({\bf x},t) = \psi^{\dag}({\bf x},t)\psi({\bf x},t),
\label{7}
\end{equation}
we obtain the continuity equation that is the local balance equation
for the number of particles
\begin{equation}
\frac{\partial n}{\partial t} + \frac{\partial j_{\mu}}{\partial x^{\mu}}=0,
\label{8}
\end{equation}
where $n({\bf x},t)=\langle \widehat{n}({\bf x},t)\rangle$ and
\begin{equation}
j_{\mu}({\bf x},t) = \langle \widehat{j}_{\mu}({\bf x},t)\rangle
=-\frac{i}{2m}\left\langle
\psi^{\dag}\frac{\partial\psi}{\partial x^{\mu}} -
\frac{\partial\psi^{\dag}}{\partial x^{\mu}}\psi
\right\rangle.
\label{9}
\end{equation}
Here the angle brackets stand for averaging with the exact
many-body density matrix. Similarly using Eq.~(\ref{6}) we derive the
equation of motion for the current, Eq.~(\ref{9}) (see, for example,
Refs.~\onlinecite{MartSchw1959,Zubarev:e})
\begin{equation}
m\frac{\partial j_{\mu}}{\partial t} + F^{\text{kin}}_{\mu} +
 F^{\text{int}}_{\mu} +
n\frac{\partial}{\partial x^{\mu}}U_{\text{ext}}=0,
\label{10}
\end{equation}
Equation (\ref{10}) has a meaning of the local force balance
equation in the fixed laboratory reference frame. Vectors
$F^{\text{kin}}_{\mu}$ and $F^{\text{int}}_{\mu}$ in
Eq.~(\ref{10}) correspond to the forces, which are related to the
kinetic and the interaction effects respectively
\begin{eqnarray}
F^{\text{kin}}_{\mu} &=&
\frac{\partial}{\partial x^{\nu}}\frac{1}{2m}\left\langle
\frac{\partial\psi^{\dag}}{\partial x^{\mu}}
\frac{\partial\psi}{\partial x^{\nu}} +
\frac{\partial\psi^{\dag}}{\partial x^{\nu}}
\frac{\partial\psi}{\partial x^{\mu}} -
\frac{\delta_{\mu\nu}}{2}\nabla^{2}\widehat{n}
\right\rangle \quad
\label{11} \\
F^{\text{int}}_{\mu} &=&
\int d{\bf x'}
\frac{\partial w(|{\bf x-x'}|)}{\partial x^{\mu}}
\rho_{2}({\bf x},{\bf x'})
\label{12}
\end{eqnarray}
In Eq.~(\ref{12}) we introduced the notation $\rho_{2}({\bf
x},{\bf x'}) = \langle \psi^{\dag}({\bf x})\psi^{\dag}({\bf
x'})\psi({\bf x'})\psi({\bf x}) \rangle$ for the two-particle
density matrix. Obviously, the last term in the left hand side in
Eq.~(\ref{10}) is the force produced by the external potential.
The kinetic force of Eq.~(\ref{11}) has a form of a divergence of
a symmetric second rank tensor. This automatically implies
vanishing integral kinetic force, $\int F^{\text{kin}}_{\mu}({\bf
x},t) d{\bf x}=0$. The Newton's third law requires that the force
$F^{\text{int}}_{\mu}$ of Eq.~(\ref{12}) should obey the same
property, which is however by far not obvious. In fact, the
possibility to represent Eq.~(\ref{12}) in a divergence form has
been a subject of a long discussion in the literature
\cite{MartSchw1959,Kugler1967,PufGil1968,Zubarev:e,NieMar1985}.
An elegant symmetric representation of the stress tensor has
been presented (unfortunately without derivation) by Puff and
Gillis in Ref.~\onlinecite{PufGil1968}. Since this representation
is of primary importance for our paper, below we give a compact
derivation of the Puff and Gillis result.

The symmetry of the function $\rho_{2}({\bf x},{\bf x'})$ with
respect to the permutation of coordinates allows us to transform
vector $F^{\text{int}}_{\mu}$, Eq.~(\ref{12}), as follows
\begin{eqnarray}\nonumber
&&F^{\text{int}}_{\mu}({\bf x}) =
\int d{\bf x'}
\frac{\partial w(|{\bf x-x'}|)}{\partial x^{\mu}}
\rho_{2}({\bf x},{\bf x'})\\
\nonumber
&=& \frac{1}{2}\int d{\bf x'}
\frac{\partial w(|{\bf x'}|)}{\partial {x'}^{\mu}}
[\rho_{2}({\bf x-x'},{\bf x}) + \rho_{2}({\bf x},{\bf x-x'})]\\
\nonumber
&=& -\frac{1}{2}\int d{\bf x'}
\frac{\partial w(|{\bf x'}|)}{\partial {x'}^{\mu}}
[\rho_{2}({\bf x+x'},{\bf x}) - \rho_{2}({\bf x},{\bf x-x'})] \\
\nonumber
&=& -\frac{1}{2}\int d{\bf x'}
\left[e^{{x'}^{\nu}\partial_{\nu}} -1 \right]
\frac{\partial w(|{\bf x'}|)}{\partial {x'}^{\mu}}
\rho_{2}({\bf x},{\bf x-x'})
\end{eqnarray}
where $\partial_{\nu}=\frac{\partial}{\partial x_{\nu}}$. Using an
obvious operator identity
$$
e^{\bf x'\nabla} - 1 = \int_{0}^{1}{\bf x'\nabla}
e^{\lambda{\bf x'\nabla}}d\lambda
$$
we arrive at the following final representation for the local force
$F^{\text{int}}_{\mu}$
\begin{equation}
F^{\text{int}}_{\mu}({\bf x}) =
\frac{\partial}{\partial x^{\nu}}W_{\mu\nu}({\bf x})
\label{13}
\end{equation}
where $W_{\mu\nu}({\bf x})$ is a stress tensor, which is
responsible for the contribution of interparticle interaction to the
force balance \cite{note1}
\begin{eqnarray} \nonumber
W_{\mu\nu}({\bf x}) = &-&\frac{1}{2}\int d{\bf x'}
\frac{{x'}^{\mu}{x'}^{\nu}}{|{\bf x'}|}
\frac{\partial w(|{\bf x'}|)}{\partial |{\bf x'}|}\\
&\times& \int_{0}^{1}
\rho_{2}({\bf x}+\lambda{\bf x'},{\bf x}-(1-\lambda){\bf x'})
d\lambda
\label{14}
\end{eqnarray}
In the next section we will show that parameter $\lambda$ in
Eq.~(\ref{14}) has a deep geometric meaning. It can be associated to
the natural parameter for a geodesic (straight line
in the present case) that connects two interacting particles (see also
Appendix~A).

Equations (\ref{11}), (\ref{13}) and (\ref{14}) show that the
net internal force, $F^{\text{kin}}_{\mu}+F^{\text{int}}_{\mu}$, is
representable in a form of divergence of a symmetric second-rank
tensor $\Pi_{\mu\nu}$. Tensor
$\Pi_{\mu\nu}$ describes local internal stresses in the fluid.
A contribution of the convective motion of particles to this tensor
is known exactly \cite{LandauVI:e}. It equals to the macroscopic momentum flow
tensor, $mnv_{\mu}v_{\nu}$,
where ${\bf v}={\bf j}/n$ is the fluid's velocity. It is
convenient to separate this contribution explicitly and rewrite
the conservation laws of Eqs.~(\ref{8}), (\ref{10}) in the
following familiar form
\begin{eqnarray}
&&D_{t}n + n\frac{\partial}{\partial x^{\mu}}v_{\mu} = 0,
\label{15}\\
&&mnD_{t}v_{\mu} + \frac{\partial}{\partial x^{\nu}}P_{\mu\nu} +
n\frac{\partial}{\partial x^{\mu}}U_{\text{ext}}=0,
\label{16}
\end{eqnarray}
where $D_{t}=\frac{\partial}{\partial t} + {\bf v}\nabla$ is the
convective derivative and $P_{\mu\nu}= T_{\mu\nu}+W_{\mu\nu}$ is
the exact stress tensor, which contains the kinetic, $T_{\mu\nu}$, and
the interaction, $W_{\mu\nu}$, contributions. The interaction stress
tensor, $W_{\mu\nu}$, is given by Eq.~(\ref{14}), while the kinetic
part, $T_{\mu\nu}$, is defined as follows
\begin{equation}
T_{\mu\nu}=\frac{1}{2m}\langle
(\widehat{q}_{\mu}\psi)^{\dag}\widehat{q}_{\nu}\psi +
(\widehat{q}_{\nu}\psi)^{\dag}\widehat{q}_{\mu}\psi -
\frac{1}{2}\delta_{\mu\nu}\nabla^{2}\widehat{n}
\rangle
\label{17}
\end{equation}
where $\widehat{\bf q}=-i\nabla - m{\bf v}$ is the operator of
``relative'' momentum which accounts for the above-mentioned separation
of the macroscopic convective motion.

Equations (\ref{15}) and (\ref{16}) form a basis for a
hydrodynamic description of a nonequilibrim many-body system.
According to the Runge-Gross mapping theorem of  TDDFT
\cite{RunGro1984} the exact many-body wave function/density matrix
(for given initial conditions) is a unique functional of velocity
${\bf v}({\bf x},t)$. Therefore the stress tensor $P_{\mu\nu}$ is
also a functional of ${\bf v}$. Hence Eqs. (\ref{15}) and
(\ref{16}) can be viewed as a formally closed system of equations
that completely determine the dynamics of collective variables
$n({\bf x},t)$ and ${\bf v}({\bf x},t)$. This dynamics is governed
by the external force, $n\partial_{\mu}U_{\text{ext}}$, and by the
force of internal stress, $\partial_{\nu}P_{\mu\nu}$. Since the
convective motion has been explicitly separated from the stress
tensor, only the relative motion of particles contributes to
$P_{\mu\nu}$. A particular form of $P_{\mu\nu}$ should be
determined from the solution of a many-body problem in a reference
frame moving with the ``center-of-mass'' velocity ${\bf v}({\bf
x},t)$. In the rest of the present paper we derive equations of
many-body dynamics in this co-moving frame and present simple
illustrative examples of their solutions.

\section{Quantum dynamics in the Lagrangian frame}

\subsection{Definition of the Lagrangian reference frame}

Co-moving or Lagrangian frame is a local noninertial reference
frame which moves with the velocity ${\bf v}({\bf x},t)$ of the
fluid. Formally the transformation to the Lagrangian frame corresponds
to a nonlinear change of variables ${\bf x}={\bf x}({\bm \xi},t)$,
which maps old coordinates ${\bf x}$ to new coordinates ${\bm
  \xi}$.  For a given velocity distribution the transformation
function, ${\bf x}({\bm \xi},t)$, is defined by the following
initial value problem 
\begin{equation}
\frac{\partial {\bf x}({\bm \xi},t)}{\partial t} =
{\bf v}({\bf x}({\bm \xi},t),t), \quad {\bf x}({\bm \xi},0)={\bm \xi}
\label{18}
\end{equation}
If the function ${\bf v}({\bf x},t)$ is continuous and satisfies the
Lipschitz condition in ${\bf x}$, there exists a unique solution to
the first order differential equation of Eq.~(\ref{18})
\cite{Korn&Korn}. Therefore, under the above conditions on the
velocity distribution, the map: ${\bf x}\to \bm\xi$ is unique and
invertible.

Physically the function ${\bf x}({\bm \xi},t)$ corresponds to the
trajectory of an infinitesimally small fluid element. Every fluid
element (and therefore every trajectory) is uniquely labeled by
the element's initial position -- the Lagrangian coordinate ${\bm
\xi}$. The inverse function ${\bm \xi}={\bm \xi}({\bf x},t)$,
which determines the transformation from the Lagrangian to the
laboratory reference frame, recovers the initial position of a
fluid element that at instant $t$ arrives at the point ${\bf x}$.
The nonlinear transformation of coordinates, ${\bf x}={\bf x}({\bm
\xi},t)$, induces a change of metric
\begin{equation}
(d{\bf x})^{2} = g_{\mu\nu}d\xi^{\mu}d\xi^{\nu}, \qquad
g_{\mu\nu}= \frac{\partial x^{\alpha}}{\partial \xi^{\mu}}
            \frac{\partial x^{\alpha}}{\partial \xi^{\nu}}
\label{19}
\end{equation}
In classical continuum mechanics the symmetric second rank tensor
$g_{\mu\nu}({\bm \xi},t)$, Eq.~(\ref{19}), is known as Green's
deformation tensor \cite{PhysAc}. This tensor is normally used to
characterize a deformed state of a system within the Lagrangian
description. The corresponding contravariant tensor, $g^{\mu\nu}$, is
defined as follows
\begin{equation}
g^{\mu\alpha}g_{\alpha\nu} = \delta^{\mu}_{\nu}, \qquad
g^{\mu\nu}= \frac{\partial \xi^{\mu}}{\partial x^{\alpha}}
            \frac{\partial \xi^{\nu}}{\partial x^{\alpha}}
\label{20}
\end{equation}
Since the deformation tensor $g_{\mu\nu}$ has a meaning of the metric
tensor in the Lagrangian $\bm\xi$-space, it should
play a key role in the description of many-body dynamics.
It is quite natural to expect that
the general equations of motion in the Lagrangian frame should reduce to
those in a space with time-dependent metric $g_{\mu\nu}({\bm
  \xi},t)$. Below we confirm this intuitive expectations by explicit
calculations.

\subsection{Equations of motion in a local noninertial reference frame}

In this subsection we derive quantum equations of motion in a
general noninertial (not necessarily Lagrangian) reference frame.
The frame is defined by its velocity ${\bf v}({\bf x},t)$, which
enters the trajectory equation of Eq.~(\ref{18}), and thus
provides a unique and invertible map: ${\bf x}\to\bm\xi$.  As a
first step in the derivation we perform a nonlinear transformation
of coordinates, ${\bf x}={\bf x}({\bm \xi},t)$, in the equation of
motion, Eq.~(\ref{6}), and in the commutation relations of
Eq.~(\ref{5}). Straightforward calculations lead to the result
\begin{widetext}
\begin{equation}
i\frac{\partial}{\partial t}\psi({\bm\xi}) = \left[
-\frac{1}{2m}\frac{1}{\sqrt{g}}\frac{\partial}{\partial\xi^{\mu}}
\sqrt{g}g^{\mu\nu}\frac{\partial}{\partial\xi^{\nu}}
+i\widetilde{v}^{\mu}({\bm\xi},t)\frac{\partial}{\partial\xi^{\mu}}
+ U_{\text{ext}}({\bm\xi},t)\right]\psi({\bm\xi})
+\int d{\bm\xi'}w(l_{\bm\xi,\bm\xi'})
\psi^{\dag}({\bm\xi'})\psi({\bm\xi'})\psi({\bm\xi})
\label{21}
\end{equation}
\end{widetext}
where
$U_{\text{ext}}({\bm\xi},t)=U_{\text{ext}}({\bf x}({\bm\xi},t),t)$,
and field operators $\psi({\bm\xi},t)$ satisfy the following
equal-time commutation relations
\begin{equation}
[\psi^{\dag}(\bm\xi),\psi(\bm\xi')]_{\pm}=
\frac{1}{\sqrt{g}}\delta({\bm\xi-\bm\xi'}).
\label{22}
\end{equation}
To shorten the notations in Eq.~(\ref{21}) we omitted the explicit
time dependence in the argument of $\psi$-operators.
The first term in the brackets in Eq.~(\ref{21}) is the Laplace
operator in a reference frame with metrics $g_{\mu\nu}$ (see, for
example, Ref.~\onlinecite{DubrovinI}), while the second term comes
from the transformation of the time derivative in Eq.~(\ref{6}). This
term is proportional to vector $\widetilde{v}^{\mu}({\bm\xi},t)$
that is the vector of velocity, transformed to a new frame:
\begin{equation}
\widetilde{v}^{\mu}({\bm\xi},t)=
\frac{\partial \xi^{\mu}}{\partial x^{\nu}}
v^{\nu}({\bf x}({\bm\xi},t),t).
\label{23}
\end{equation}
The interparticle distance, $l_{\bm\xi,\bm\xi'}$, in the argument of the
interaction potential in Eq.~(\ref{21}) equals to a length of
geodesic that connects points ${\bm\xi}$ and ${\bm\xi'}$.
Geodesic, ${\bf z}_{\bm\xi,\bm\xi'}(\lambda)$, parameterized by a
parameter $\lambda$ ($0<\lambda<1$), is a solution to the following
equation \cite{DubrovinI}
\begin{equation}
\ddot{z}^{\mu}(\lambda) + \Gamma^{\mu}_{\alpha\beta}({\bf z})
\dot{z}^{\alpha}(\lambda)\dot{z}^{\beta}(\lambda)=0,
\label{24}
\end{equation}
where $\dot{\bf z}=\partial{\bf z}/\partial\lambda$, and
$\Gamma^{\mu}_{\alpha\beta}$ is affine connection
\begin{equation}
\Gamma^{\mu}_{\alpha\beta} = \frac{1}{2}g^{\mu\nu}\left(
\frac{\partial g_{\nu\alpha}}{\partial \xi^{\beta}} +
\frac{\partial g_{\nu\beta}}{\partial \xi^{\alpha}} -
\frac{\partial g_{\alpha\beta}}{\partial \xi^{\nu}}
\right)
\label{25}
\end{equation}
Equation (\ref{24}) should be solved with boundary conditions
${\bf z}(0)={\bm\xi}$, ${\bf z}(1)={\bm\xi}'$. It is convenient to
parameterize geodesics by a natural parameter, which is chosen in
such a way that an absolute value of the ``velocity'', $|\dot{\bf
z}| =\sqrt{g_{\mu\nu}\dot{z}^{\mu}\dot{z}^{\nu}}$, becomes
independent of $\lambda$ along the curve ${\bf z}(\lambda)$. For
this parameterization the length $l_{\bm\xi,\bm\xi'}$, which enters
Eq.~(\ref{21}), simply equals to $|\dot{\bf z}|$ at any point on
the geodesic
\begin{equation}
l_{\bm\xi,\bm\xi'} = \int_{0}^{1}
\sqrt{g_{\mu\nu}({\bf z})
\dot{z}^{\mu}(\lambda)\dot{z}^{\nu}(\lambda)}d\lambda =
\sqrt{g_{\mu\nu}\dot{z}^{\mu}\dot{z}^{\nu}}
\label{26}
\end{equation}

Equation (\ref{21}) is the equation of motion for the operator
$\psi(\bm\xi,t)=\psi({\bf x}({\bm\xi},t),t)$.
Due to the Jacobian factor $1/\sqrt{g}$ in the commutation relations of
Eq.~(\ref{22}), the quantity $\psi(\bm\xi,t)$ can not be interpreted as
an operator for annihilation of a particle in a given point of
${\bm\xi}$-space. In particular, the operator
$\widehat{n}(\bm\xi)=\psi^{\dag}(\bm\xi)\psi(\bm\xi)$ does not
correspond to the density operator in the new frame. By
definition the density is a number of particles per unit
volume that is changed under a volume non-preserving coordinate
transformation. Therefore it is natural to define the physical field
operators and the density operator as follows
\begin{eqnarray}
\widetilde{\psi}(\bm\xi)&=&g^{\frac{1}{4}}\psi(\bm\xi), \quad
\widetilde{\psi}^{\dag}(\bm\xi)=g^{\frac{1}{4}}\psi^{\dag}(\bm\xi),
\label{27}\\
\widehat{\widetilde{n}}(\bm\xi) &=&
\widetilde{\psi}^{\dag}(\bm\xi)\widetilde{\psi}(\bm\xi)
=\sqrt{g}\psi^{\dag}(\bm\xi)\psi(\bm\xi),
\label{28}
\end{eqnarray}
which automatically accounts for the proper change of a unit
volume in the deformed reference frame. Obviously the redefined field operators
$\widetilde{\psi}(\bm\xi)$ satisfy the common commutations relations
\begin{equation}
[\widetilde{\psi}^{\dag}(\bm\xi),\widetilde{\psi}(\bm\xi')]_{\pm}=
\delta({\bm\xi-\bm\xi'}).
\label{29}
\end{equation}
The renormalization of $\psi$-operators,
Eq.~(\ref{27}), is equivalent to the corresponding multiplicative
redefinition of the many-body wave function. This redefinition is aimed to
preserve the common probabilistic interpretation and the standard form
of the normalization conditions in the new reference frame (similar
arguments were suggested by Podolsky \cite{Podolsky1928} in
early days of quantum mechanics).

Let us show that the renormalization of field operators,
Eq.~(\ref{27}), also simplifies the form of the equations of motion.
First we note that the differential operator in the right hand
side in Eq.~(\ref{21}) (first two terms in the square brackets) can be
rearranged as follows
\begin{eqnarray}\nonumber
&-&\frac{1}{2m}\frac{1}{\sqrt{g}}\frac{\partial}{\partial\xi^{\mu}}
\sqrt{g}g^{\mu\nu}\frac{\partial}{\partial\xi^{\nu}}
+i\widetilde{v}^{\mu}\frac{\partial}{\partial\xi^{\mu}} \\
&=& \frac{1}{\sqrt{g}}
\frac{\widehat{K}_{\mu}\sqrt{g}\widehat{K}^{\mu}}{2m} -
m\frac{\widetilde{v}_{\mu}\widetilde{v}^{\mu}}{2}
-i\frac{1}{2\sqrt{g}}\left(\frac{\partial}{\partial\xi^{\mu}}
\sqrt{g}\widetilde{v}^{\mu}\right) \label{30}
\end{eqnarray}
where we introduced an operator of ``kinematic'' momentum in
the noninertial reference frame:
\begin{equation}
\widehat{K}_{\mu} =
 -i\frac{\partial}{\partial\xi^{\mu}} - m\widetilde{v}_{\mu}.
\label{31}
\end{equation}
(Raising and lowering of tensor indices are performed according to the
standard rules, i.e.
$\widetilde{v}_{\mu}=g_{\mu\nu}\widetilde{v}^{\nu}$ or
$\widehat{K}^{\mu}=g^{\mu\nu}\widehat{K}_{\nu}$.)

Using the equation of trajectory ${\bf x}({\bm\xi},t)$,
Eq.~(\ref{18}), and the definition of metric tensor $g_{\mu\nu}$,
Eq.~(\ref{19}), one can prove the following identity
\begin{equation}
g^{-\frac{1}{4}}\frac{\partial g^{\frac{1}{4}}}{\partial t}=
\frac{1}{4}\frac{\partial \ln g}{\partial t} =
\frac{1}{2\sqrt{g}}\left(\frac{\partial}{\partial\xi^{\mu}}
\sqrt{g}\widetilde{v}^{\mu}\right)
\label{32}
\end{equation}
The quantity in the right hand side in Eq.~(\ref{32}) coincides with the last
term in the right hand side in Eq.~(\ref{30}). Hence the sum of
the corresponding term in the equation of motion, Eq.~(\ref{21}), and
of the time derivative of $\psi$ reduces to the following compact form
\begin{equation}
\frac{\partial\psi}{\partial t} +
\frac{1}{2\sqrt{g}}\left(\frac{\partial}{\partial\xi^{\mu}}
\sqrt{g}\widetilde{v}^{\mu}\right)\psi
= g^{-\frac{1}{4}}\frac{\partial g^{\frac{1}{4}}\psi}{\partial t}
= g^{-\frac{1}{4}}\frac{\partial \widetilde{\psi}}{\partial t}
\label{32a}
\end{equation}
Substituting Eq.~(\ref{30}) into Eq.~(\ref{21}) and using
Eq.~(\ref{32a}), we obtain the final equation of motion for the
renormalized field operator $\widetilde{\psi}({\bm\xi},t)$
\begin{eqnarray}\nonumber
i\frac{\partial \widetilde{\psi}({\bm\xi})}{\partial t}&=& \left[
g^{-\frac{1}{4}}\frac{
\widehat{K}_{\mu}\sqrt{g}\widehat{K}^{\mu}}{2m}g^{-\frac{1}{4}}
+U_{\text{ext}} -
m\frac{\widetilde{v}_{\mu}\widetilde{v}^{\mu}}{2}
\right]\widetilde{\psi}({\bm\xi})\\
&+& \int d{\bm\xi'}w(l_{\bm\xi,\bm\xi'})
\widetilde{\psi}^{\dag}({\bm\xi'})\widetilde{\psi}({\bm\xi'})
\widetilde{\psi}({\bm\xi})
\label{33}
\end{eqnarray}
Equation~(\ref{33}) allows us to recover a form of the transformed Hamiltonian
$\widetilde{H}[\widetilde{\psi}^{\dag},\widetilde{\psi}]$, which,
together with the commutation relations of Eq.~(\ref{29}),
determines the dynamics of the system
\begin{eqnarray}
\widetilde{H} &=&
\widehat{\widetilde{T}}+\widehat{\widetilde{W}}+\widehat{\widetilde{U}},
\label{34}\\
\widehat{\widetilde{T}}&=& \int d{\bm\xi}\sqrt{g}
\left(\widehat{K}_{\mu}g^{-\frac{1}{4}}\widetilde{\psi}\right)^{\dag}
\frac{g^{\mu\nu}}{2m}
\left(\widehat{K}_{\nu}g^{-\frac{1}{4}}\widetilde{\psi}\right),
\label{35}\\
\widehat{\widetilde{W}}&=& \frac{1}{2}\int
d{\bm\xi}d{\bm\xi'}w(l_{\bm\xi,\bm\xi'})\widetilde{\psi}^{\dag}({\bm\xi})
\widetilde{\psi}^{\dag}({\bm\xi'})\widetilde{\psi}({\bm\xi'})
\widetilde{\psi}({\bm\xi}),
\label{36}\\
\widehat{\widetilde{U}}&=& \int d{\bm\xi}\left(U_{\text{ext}}-
m\frac{\widetilde{v}_{\mu}\widetilde{v}^{\mu}}{2}\right)\widetilde{\psi}^{\dag}\widetilde{\psi}
\label{37}
\end{eqnarray}
Equations (\ref{33})--(\ref{37}) represent the main results of this
subsection. Equation (\ref{33}) is the Heisenberg equation of motion
for the physical field operator, while Eqs.~(\ref{34})--(\ref{37})
establish the rules for the transformation of the many-body
Hamiltonian to an arbitrary local noninertial reference frame.

Formally the Hamiltonian of Eqs.~(\ref{34})--(\ref{37}) describes a
system of quantum particles in the presence of an effective vector
potential $m\widetilde{\bf v}({\bm\xi},t)$ and an additional
effective scalar potential $m{\widetilde{\bf v}}^{2}/2$. The
particles live in a space with the time-dependent metric
$g_{\mu\nu}({\bm\xi},t)$ and interact via pairwise potential which
depends on the length of geodesic connecting a pair of particles.
Additional "potentials" and a nontrivial metric tensor are
responsible for generalized inertia forces exerted on a particle
in a general noninertial reference frame. To get a transparent
physical understanding of these forces it is instructive to look
on dynamics in the semiclassical approximation. Since the most
important inertial contributions enter only quadratic parts of the
Hamiltonian (Eqs.~(\ref{35}) and (\ref{37})), we neglect for a
moment the interaction, and consider an equation of motion for the
Wigner function
$$
\widetilde{f}_{\bf p}({\bm\xi},t)=\int e^{-i{\bf p}{\bm\eta}}
\langle \widetilde{\psi}^{\dag}({\bm\xi}+\frac{\bm\eta}{2},t)
\widetilde{\psi}({\bm\xi}-\frac{\bm\eta}{2},t) \rangle d{\bm\xi}
$$
in a gas of noninteracting particles. In the semiclassical limit
the Wigner function satisfies the following kinetic equation
\begin{equation}
\frac{\partial \widetilde{f}_{\bf p}}{\partial t} + \frac{\partial
\widetilde{H}({\bf p},{\bm\xi})}{\partial {\bf p}} \frac{\partial
\widetilde{f}_{\bf p}}{\partial {\bm\xi}} - \frac{\partial
\widetilde{H}({\bf p},{\bm\xi})}{\partial {\bm\xi}} \frac{\partial
\widetilde{f}_{\bf p}}{\partial {\bf p}}=0,
\label{38}
\end{equation}
where $\widetilde{H}({\bf p},{\bm\xi})$ is the semiclassical
Hamiltonian function, which corresponds the noninteracting part of
Eq.~(\ref{34}):
\begin{eqnarray}\nonumber
\widetilde{H}({\bf p},{\bm\xi}) &=& \frac{g^{\mu\nu}}{2m}
(p_{\mu}-m\widetilde{v}_{\mu})(p_{\nu}-m\widetilde{v}_{\nu})\\
&+& U_{\text{ext}} -
m\frac{\widetilde{v}_{\mu}\widetilde{v}^{\mu}}{2}.
\label{39}
\end{eqnarray}
Substituting Eq.~(\ref{39}) into Eq.~(\ref{38}) we get the result
\begin{eqnarray}\nonumber
\frac{\partial \widetilde{f}_{\bf p}}{\partial t} &+&
\frac{g^{\mu\nu}}{m}(p_{\mu}-m\widetilde{v}_{\mu})\frac{\partial
\widetilde{f}_{\bf p}}{\partial \xi^{\nu}}\\
&-& \left(\frac{\partial g^{\alpha\beta}}{\partial \xi^{\nu}}
\frac{p_{\alpha}p_{\beta}}{2m} -  \frac{\partial
\widetilde{v}^{\alpha}}{\partial \xi^{\nu}}p_{\alpha} +
\frac{\partial U_{\text{ext}}}{\partial \xi^{\nu}}\right)
\frac{\partial \widetilde{f}_{\bf p}}{\partial p_{\nu}} = 0. \quad
 \label{40}
\end{eqnarray}
Inertia forces do not explicitly show up in Eq.~(\ref{40}). The
reason is that Eq.~(\ref{40}) is the equation for the function
$\widetilde{f}_{\bf p}$ which depends on the canonical momentum ${\bf
p}$. The physical velocity of a particle in the new reference
frame is proportional to the kinematic momentum ${\bf K}={\bf p} -
m\widetilde{\bf v}$ (i.e. $\partial\widetilde{H}/\partial p_{\mu} =
K^{\mu}/m$). Therefore it is more natural physically to consider
${\bf K}$ as an independent variable in the kinetic equation. The
distribution function of the kinematic momentum, $\widetilde{f}'_{\bf
K}({\bm\xi},t)$, can be introduced as follows
\begin{equation}
\widetilde{f}'_{\bf K}({\bm\xi},t) = \widetilde{f}_{{\bf K} +
m\widetilde{\bf v}}({\bm\xi},t).
\label{41}
\end{equation}
Performing the corresponding change of
variables in Eq.~(\ref{40}) we obtain the final semiclassical
equation of motion for the distribution function $\widetilde{f}'_{\bf
K}({\bm\xi},t)$ in the local noninertial reference frame
\begin{eqnarray}\nonumber
\frac{\partial \widetilde{f}'_{\bf K}}{\partial t} &+&
\frac{K^{\nu}}{m}\frac{\partial \widetilde{f}'_{\bf K}}{\partial
\xi^{\nu}} - \Big[m\frac{\partial \widetilde{v}_{\nu}}{\partial t}
+ K^{\mu}\widetilde{F}_{\mu\nu}
 - \frac{\partial g_{\alpha\beta}}{\partial \xi^{\nu}}
\frac{K^{\alpha}K^{\beta}}{2m} \\
&+& \frac{\partial}{\partial\xi^{\nu}}\left(U_{\text{ext}} -
m\frac{\widetilde{v}_{\mu}\widetilde{v}^{\mu}}{2}\right)\Big]
\frac{\partial \widetilde{f}'_{\bf K}}{\partial K_{\nu}} = 0,
 \label{42}
\end{eqnarray}
where a skew-symmetric second rank tensor $\widetilde{F}_{\mu\nu}$
is defined as follows
\begin{equation}
\widetilde{F}_{\mu\nu} =
\frac{\partial\widetilde{v}_{\mu}}{\partial\xi^{\nu}} -
\frac{\partial\widetilde{v}_{\nu}}{\partial\xi^{\mu}}.
\label{43}
\end{equation}
Since tensor $\widetilde{F}_{\mu\nu}$ vanishes for an
irrotational flow, we name it the
vorticity tensor \cite{note2}. In the next section Eq.~(\ref{42}) will be
applied to the derivation of generalized collisionless
hydrodynamics \cite{TokPPRB1999,TokPPRB2000}.

The expression in the square brackets in Eq.~(\ref{42}) contains
all inertia forces. These are all the terms except for the
external force, $\partial_{\nu}U_{\text{ext}}$.  The first term in
the square brackets is the linear acceleration force, while the
last term is related to the kinetic energy of a moving frame. In a
particular case of a homogeneously rotating frame the last term is
responsible for the centrifugal force. The second and the third
terms in the square brackets correspond to inertia forces that
depend on a velocity of a particular particle. The second term is
the classical Coriolis force. This force is proportional to the
skew-symmetric vorticity tensor, which defines a local angular
velocity of the reference frame. The third, bilinear in particle's
momentum term is less common. The corresponding inertia force
makes a free particle to move along a geodesic in a local
noninertial frame. Indeed, the third term in the square brackets
in Eq.~(\ref{42}) can be rewritten as follows
\begin{equation}
\frac{1}{2m}\frac{\partial g_{\alpha\beta}}{\partial \xi^{\nu}}
K^{\alpha}K^{\beta} = \frac{1}{m}
g_{\nu\mu}\Gamma^{\mu}_{\alpha\beta}K^{\alpha}K^{\beta},
\label{44}
\end{equation}
where we have used Eq.~(\ref{25}), which relates the affine connection
$\Gamma^{\mu}_{\alpha\beta}$ to the metric tensor $g_{\mu\nu}$.
The right hand side in Eq.~(\ref{44}) is easily recognized as a
covariant component of the force in the
equation of geodesic (see, for example, Eq.~(\ref{24})).

\subsection{Conserving quantities and balance equations}

\subsubsection{The continuity equation}
The first problem we address in this subsection is a proper
definition of the current operator,
$\widehat{\widetilde{j}^{\mu}}$, in a general noninertial
reference frame. The easiest way to establish a form of
$\widehat{\widetilde{j}^{\mu}}$ is to derive the equation of
motion for the density operator $\widehat{\widetilde{n}}(\bm\xi,t)
= \widetilde{\psi}^{\dag}(\bm\xi,t)\widetilde{\psi}(\bm\xi,t)$.
Using Eq.~(\ref{33}) to compute the time derivative of the
density operator we find that the desired equation indeed reduces
to the common form of the continuity equation,
\begin{equation}
\frac{\partial \widehat{\widetilde{n}}}{\partial t} +
\frac{\partial \widehat{\widetilde{j}^{\mu}}}{\partial \xi^{\mu}} = 0,
\label{45}
\end{equation}
if we define the current operator, 
$\widehat{\widetilde{j}^{\mu}}(\bm\xi,t)$, as follows
\begin{equation}
\widehat{\widetilde{j}^{\mu}} = g^{\mu\nu}\left[\frac{-i}{2m}\left(
 \widetilde{\psi}^{\dag}
\frac{\partial\widetilde{\psi}}{\partial\xi^{\nu}} -
\frac{\partial\widetilde{\psi}^{\dag}}{\partial\xi^{\nu}}
\widetilde{\psi}\right) -
\widetilde{v}_{\nu}\widetilde{\psi}^{\dag}\widetilde{\psi}
 \right].
\label{46}
\end{equation}
The standard form of the continuity equation, Eq.~(\ref{45}), should be
considered as one more justification for the redefinition of field
operators, Eq.~(\ref{27}). We would like to outline a very natural
form of the current operator, Eq.~(\ref{46}). Despite the presence of
the Jacobian factors ($\sqrt{g}$ or $g^{1/4}$) in the Hamiltonian,
they completely vanish in Eq.~(\ref{46}) (as well as in the definition
of the density operator of Eq.~(\ref{28})).

From this point we restrict ourselves to the Lagrangian frame, which is
the local reference frame, moving with the velocity ${\bf v}$ of the fluid. 
In this special case the continuity equation admits a
very simple solution. Let us calculate the expectation value of the
current operator $\widehat{\widetilde{j}^{\mu}}$, Eq.~(\ref{46}).
This can be done, for example, by transforming the right hand side
in Eq.~(\ref{46}) back to the laboratory frame, and by using
Eq.~(\ref{9}) together with the definition of the velocity, ${\bf
v}={\bf j}/n$. The result takes an extremely simple form
\begin{equation}
\widetilde{j}^{\mu}(\bm\xi,t) = \langle
\widehat{\widetilde{j}^{\mu}}(\bm\xi,t)\rangle = 0.
\label{47}
\end{equation}
Thus the current density is exactly zero in every point of the
Lagrangian $\bm\xi$-space. This is of course not surprising, since
an observer in the co-moving frame should not see any
current. Combining Eq.~(\ref{47}) and the continuity
equation of Eq.~(\ref{45}) we arrive at the conclusion that the
density $\widetilde{n}(\bm\xi,t)$ is independent of time
\begin{equation}
\widetilde{n}(\bm\xi,t) = \widetilde{n}(\bm\xi,0) = n_{0}(\bm\xi),
\label{48}
\end{equation}
where $n_{0}({\bf x})$ is the initial density distribution.
Therefore in the Lagrangian frame not only the number of particles $N$
is an integral of motion, but the density itself is also a
conserving quantity. Evolution of the density in the laboratory
frame can be calculated with the following formula (see
Eq.~(\ref{28}))
\begin{equation}
n({\bf x},t) =
\frac{\widetilde{n}(\bm\xi({\bf x},t),t)}{\sqrt{g(\bm\xi({\bf x},t),t)}}
 = \frac{n_{0}(\bm\xi({\bf x},t))}{\sqrt{g(\bm\xi({\bf x},t),t)}}
\label{49}
\end{equation}
Equation (\ref{49}) is, in fact, the explicit solution to the
continuity equation of Eq.~(\ref{8}), which defines the density
$n({\bf x},t)$ as a functional of velocity ${\bf v}({\bf x},t)$.

Equations (\ref{47}) and (\ref{48}) demonstrate the main advantage
of the Lagrangian frame for the description of many-body dynamics. In
this very special reference frame the inertia forces are adjusted
to get exactly zero current density and therefore to keep the
density of particles fixed during the whole evolution of the
system. Equation (\ref{47}) can be used to construct a complete
many-body theory in the co-moving frame. The frame's velocity
${\bf v}$ enters the equation of motion, Eq.~(\ref{33}), as an
external parameter. Imposing the local ``gauge'' condition of Eq.
~(\ref{47}) we specify the reference frame and thus get the complete
theory with all quantities defined by the initial conditions.

\subsubsection{Local force balance in the Lagrangian frame}
Let us turn to the local momentum conservation law. In the
laboratory reference frame it is given by Eq.~(\ref{16}) (or,
equivalently, by Eq.~(\ref{10})). Since in the Lagrangian frame the
current density is zero, the local momentum conservation law
should reduce to the zero force condition -- the inertia forces should
exactly compensate the external force and the force of internal
stresses. Below we derive an explicit form of this balance
equation by the direct transformation of Eq.~(\ref{16}) to
the Lagrangian coordinates $\bm\xi$. First we express the vector of
velocity ${\bf v}$ and the stress tensor $P_{\mu\nu}$ in terms of
the corresponding quantities, $\widetilde{\bf v}$ and
$\widetilde{P}^{\mu}_{\nu}$, in the Lagrangian frame
\begin{eqnarray}
v_{\mu} &=& \frac{\partial\xi^{\beta}}{\partial x^{\mu}}
\widetilde{v}_{\beta}({\bm\xi},t),
\label{50} \\
P_{\mu\nu} &=& \frac{\partial\xi^{\beta}}{\partial x^{\mu}}
             \frac{\partial x^{\nu}}{\partial\xi^{\gamma}}
 \widetilde{P}^{\gamma}_{\beta}({\bm\xi},t).
\label{51}
\end{eqnarray}
Equation (\ref{50}) follows the definition of $\widetilde{v}^{\mu}$,
Eq.~(\ref{23}), while in Eq.~(\ref{51}) we adopted the standard
tensor transformation rules \cite{DubrovinI}. Substituting
Eqs.~(\ref{50}) and (\ref{51}) into Eq.~(\ref{16}), transforming the
derivatives, and multiplying the result by
$\partial x^{\mu}/\partial\xi^{\alpha}$, we obtain the following
equation
\begin{eqnarray}\nonumber
mn\frac{\partial x^{\mu}}{\partial\xi^{\alpha}}
\frac{\partial}{\partial t}\left(
\frac{\partial\xi^{\beta}}{\partial x^{\mu}}
\widetilde{v}_{\beta} \right)
&+&\frac{\partial x^{\mu}}{\partial\xi^{\alpha}}
\frac{\partial\xi^{\delta}}{\partial x^{\nu}}
\frac{\partial}{\partial\xi^{\delta}}\left(
\frac{\partial\xi^{\beta}}{\partial x^{\mu}}
\frac{\partial x^{\nu}}{\partial\xi^{\gamma}}
\widetilde{P}^{\gamma}_{\beta}\right)  \\
&+&n\frac{\partial U_{\text{ext}}}{\partial\xi^{\alpha}} = 0
\label{52}
\end{eqnarray}
The first term in Eq.~(\ref{52}) can be further simplified as follows
\begin{equation}
\frac{\partial x^{\mu}}{\partial\xi^{\alpha}}
\frac{\partial}{\partial t}
\frac{\partial\xi^{\beta}}{\partial x^{\mu}}
\widetilde{v}_{\beta}
= \frac{\partial\widetilde{v}_{\alpha}}{\partial t}
- \widetilde{v}_{\beta}\frac{\partial\xi^{\beta}}{\partial x^{\mu}}
\frac{\partial v^{\mu}}{\partial\xi^{\alpha}}
= \frac{\partial\widetilde{v}_{\alpha}}{\partial t}
- \frac{1}{2}\frac{\partial
\widetilde{v}_{\beta}\widetilde{v}^{\beta}}{\partial\xi^{\alpha}}
\label{53}
\end{equation}
In the sequence of transformations in Eq.~(\ref{53}) we used an obvious
identity $\frac{\partial x^{\mu}}{\partial\xi^{\alpha}}
\frac{\partial\xi^{\beta}}{\partial x^{\mu}} =
\delta_{\alpha}^{\beta}$, the trajectory equation of Eq.~(\ref{18}), and
the definition of $\widetilde{v}^{\mu}$,
Eq.~(\ref{23}). A similar simplification of the second term in
Eq.~(\ref{52}) is even more straightforward. One only needs to apply
the chain rule for the calculation of the spatial derivative and
take into account the following explicit representation for affine
connection (see, for example, Ref.~\onlinecite{DubrovinI}),
\begin{equation}
\Gamma_{\alpha\beta}^{\mu} =
\frac{\partial\xi^{\mu}}{\partial x^{\gamma}}
\frac{\partial^{2} x^{\gamma}}{\partial\xi^{\alpha}\partial\xi^{\beta}}.
\label{54}
\end{equation}
As a result we get a very natural expression for the second term in
Eq.~(\ref{52})
\begin{equation}
\frac{\partial x^{\mu}}{\partial\xi^{\alpha}}
\frac{\partial\xi^{\delta}}{\partial x^{\nu}}
\frac{\partial}{\partial\xi^{\delta}}\left(
\frac{\partial\xi^{\beta}}{\partial x^{\mu}}
\frac{\partial x^{\nu}}{\partial\xi^{\gamma}}
\widetilde{P}^{\gamma}_{\beta}\right)
= \widetilde{P}^{\beta}_{\alpha ; \beta}
\label{55}
\end{equation}
where the semicolon is used to denote the covariant
derivative. The covariant divergence of the stress tensor in
Eq.~(\ref{55}) is defined as follows \cite{DubrovinI,LandauII:e}
\begin{eqnarray}\nonumber
\widetilde{P}^{\nu}_{\mu ; \nu} &=&
\frac{\partial\widetilde{P}^{\nu}_{\mu}}{\partial\xi^{\nu}}
+ \Gamma^{\nu}_{\nu\alpha}\widetilde{P}^{\alpha}_{\mu}
- \Gamma^{\nu}_{\mu\alpha}\widetilde{P}^{\alpha}_{\nu} \\
&=& \frac{1}{\sqrt{g}}
\frac{\partial\sqrt{g}\widetilde{P}^{\nu}_{\mu}}{\partial\xi^{\nu}}
- \frac{1}{2}\frac{\partial g_{\alpha\beta}}{\partial\xi^{\mu}}
\widetilde{P}^{\alpha\beta}
\label{56}
\end{eqnarray}
Substitution of Eqs.~(\ref{53}) and (\ref{55}) into Eq.~(\ref{52})
leads to the final form of the force balance equation in the Lagrangian
frame
\begin{equation}
\widetilde{n}\left[
m\frac{\partial\widetilde{v}_{\mu}}{\partial t}
+ \frac{\partial}{\partial\xi^{\mu}}\left(
U_{\text{ext}}
- m\frac{\widetilde{v}_{\nu}\widetilde{v}^{\nu}}{2}\right)
\right]
+ \sqrt{g}\widetilde{P}^{\nu}_{\mu ; \nu} = 0.
\label{57}
\end{equation}
A direct comparison of the force term in the kinetic equation of Eq.~(\ref{42}) 
and the term in
the square brackets in Eq.~(\ref{57}) shows that the later is exactly the
sum of the external force and two inertia forces that are independent of
particle's momentum. These three forces are balanced by the force of
internal stresses (the second term in
Eq.~(\ref{57})). The net force, exerted on every fluid element in the
Lagrangian space, is zero, which results in a zero current density and
a stationary particles' density distribution. It should be noted that
the rest of inertia forces (those, which are different for different particles
in a fluid element) implicitly present in the kinetic part of the
stress tensor $\widetilde{P}_{\mu\nu}$. To see this more clearly we
need to derive an explicit microscopic representation for this tensor.

\subsubsection{Stress tensor in the Lagrangian frame}
In general both kinetic, $\widetilde{T}_{\mu\nu}$, and interaction,
$\widetilde{W}_{\mu\nu}$, contributions to the total stress tensor,
$\widetilde{P}_{\mu\nu}=\widetilde{T}_{\mu\nu}+\widetilde{W}_{\mu\nu}$,
can be found using the common transformations rules
\cite{DubrovinI}
\begin{eqnarray}
\widetilde{T}_{\mu\nu}(\bm\xi,t) &=&
\frac{\partial x^{\alpha}}{\partial\xi^{\mu}}
\frac{\partial x^{\beta}}{\partial\xi^{\nu}}
T_{\alpha\beta}({\bf x}(\bm\xi,t),t),
\label{58} \\
\widetilde{W}_{\mu\nu}(\bm\xi,t) &=&
\frac{\partial x^{\alpha}}{\partial\xi^{\mu}}
\frac{\partial x^{\beta}}{\partial\xi^{\nu}}
W_{\alpha\beta}({\bf x}(\bm\xi,t),t).
\label{59}
\end{eqnarray}
Here stress tensors, $T_{\alpha\beta}({\bf x},t)$ and
$W_{\alpha\beta}({\bf x},t)$, in the laboratory reference frame are
given by Eqs.~(\ref{17}) and (\ref{14}) respectively. We shall
however follow another route, which takes full advantage of
geometric ideas we develop in this paper. The transformed many-body
Hamiltonian of Eqs.~(\ref{34})--(\ref{37}), is an explicit functional of
the metric tensor $g_{\mu\nu}$. Therefore we can find the required
stress tensor by computing the variational derivative of the
energy with respect to the metric tensor. More precisely, we make
use of the fact that under a small variation of the metric, the
variations of the kinetic energy, Eq.~(\ref{35}), and of the energy
of interparticle interaction, Eq.~(\ref{36}), are related to
tensors $\widetilde{T}_{\mu\nu}$ and $\widetilde{W}_{\mu\nu}$
respectively
\begin{eqnarray}
\delta\langle\widehat{\widetilde T}\rangle &=&
\int d{\bm\xi}\frac{\sqrt{g}}{2}\delta g^{\mu\nu}\widetilde{T}_{\mu\nu}
 = -
\int d{\bm\xi}\frac{\sqrt{g}}{2}\delta g_{\mu\nu}\widetilde{T}^{\mu\nu}
\label{60} \\
\delta\langle\widehat{\widetilde W}\rangle &=&
\int d{\bm\xi}\frac{\sqrt{g}}{2}\delta g^{\mu\nu}\widetilde{W}_{\mu\nu}
 = -
\int d{\bm\xi}\frac{\sqrt{g}}{2}\delta g_{\mu\nu}\widetilde{W}^{\mu\nu}
\label{61}
\end{eqnarray}
Such a definition of the stress tensors is closely related to the
common definition of the energy-momentum tensor in general
relativity (see, for example, Ref.~\onlinecite{LandauII:e}). The
main advantage of this definition is that it automatically gives a
symmetric form of the stress/energy-momentum tensors. Recently a
very similar approach has been used to derive a microscopic
expression for the stress tensor in the equilibrium quantum
many-body system within the local density approximation
\cite{RogRap2002}. The paper of Ref.~\onlinecite{RogRap2002} also
contains a general discussion of the above geometric definition of
the stress tensors in the context of nonrelativistic quantum
mechanics.

The variation of the Hamiltonian should be taken at constant
$\widetilde{\psi}$-variables (since they satisfy the equations of
motion) \cite{LandauII:e,RogRap2002} and at constant velocity
$\widetilde{\bf v}$. The kinetic energy operator
$\widehat{\widetilde T}$ of (\ref{35}) contains the metric tensor
only in a form of $g^{\mu\nu}$, $\sqrt{g}$ and $g^{-\frac{1}{4}}$.
Noting that
$$
\delta\sqrt{g} =
 - \frac{1}{2}\sqrt{g}g_{\mu\nu}\delta g^{\mu\nu},
\qquad
\delta g^{-\frac{1}{4}} =
   \frac{1}{4}g^{-\frac{1}{4}}g_{\mu\nu}\delta g^{\mu\nu},
$$
we can easily compute the variation of $\widehat{\widetilde T}$ and
then identify the kinetic stress tensor using Eq.~(\ref{60}). The
result of the calculations takes the form
\begin{widetext}
\begin{equation}
\widetilde{T}_{\mu\nu}(\bm\xi,t) = \frac{1}{2m}\left\langle
\left(\widehat{K}_{\mu}g^{-\frac{1}{4}}\widetilde{\psi}\right)^{\dag}
\left(\widehat{K}_{\nu}g^{-\frac{1}{4}}\widetilde{\psi}\right) +
\left(\widehat{K}_{\nu}g^{-\frac{1}{4}}\widetilde{\psi}\right)^{\dag}
\left(\widehat{K}_{\mu}g^{-\frac{1}{4}}\widetilde{\psi}\right) -
\frac{1}{2}g_{\mu\nu}\frac{1}{\sqrt{g}}
\frac{\partial}{\partial\xi^{\alpha}}\sqrt{g}g^{\alpha\beta}
\frac{\partial}{\partial\xi^{\beta}}
\frac{\widetilde{\psi}^{\dag}\widetilde{\psi}}{\sqrt{g}}\right\rangle
\label{62}
\end{equation}
\end{widetext}
If we evaluate the right hand side in Eq.~(\ref{62}) for Euclidian
metric, $g_{\mu\nu}=\delta_{\mu\nu}$, we immediately recover
Eq.~(\ref{17}). Therefore the commonly used symmetric form of the kinetic
stress tensor $T_{\mu\nu}$, Eq.~(\ref{17}), is in exact correspondence with
the geometric definition of Eq.~(\ref{60}).

Calculation of the variation $\delta\widehat{\widetilde W}$,
Eq.~(\ref{61}), is a
little bit more involved. Since the interaction Hamiltonian of Eq.~(\ref{36})
depends on $g_{\mu\nu}$ only via the length of geodesic, we have
\begin{equation}
\delta\langle\widehat{\widetilde W}\rangle = \frac{1}{2}\int
d{\bm\eta}d{\bm\eta'}\delta l_{\bm\eta,\bm\eta'}
\frac{\partial w(l_{\bm\eta,\bm\eta'})}{\partial l_{\bm\eta,\bm\eta'}}
\widetilde{\rho}_{2}(\bm\eta,\bm\eta')
\label{63}
\end{equation}
where $\widetilde{\rho}_{2}(\bm\eta,\bm\eta') = \langle
\widetilde{\psi}^{\dag}({\bm\eta})\widetilde{\psi}^{\dag}({\bm\eta'})
\widetilde{\psi}({\bm\eta'})\widetilde{\psi}({\bm\eta})\rangle$.
The next step is to compute the variation of the functional
$l_{\bm\eta,\bm\eta'}[g_{\mu\nu}]$, Eq.~(\ref{26}). Let $\lambda$
in Eq.~(\ref{26}) be a natural parameter for a geodesic in the
space with ``unperturbed'' metric $g_{\mu\nu}$ (not the ``full''
metric $g_{\mu\nu}+\delta g_{\mu\nu}$). For this parameterization
the variation of $l_{\bm\eta,\bm\eta'}[g_{\mu\nu}]$ takes the form
\begin{eqnarray}\nonumber
&&\delta l_{\bm\eta,\bm\eta'} = \frac{1}{2l_{\bm\eta,\bm\eta'}}
\int_{0}^{1}d\lambda
\delta g_{\mu\nu}({\bf z}(\lambda))
\dot{z}^{\mu}(\lambda)\dot{z}^{\nu}(\lambda) \\
&=& \int d{\bm\xi}\int_{0}^{1}d\lambda
\delta(\bm\xi - {\bf z}(\lambda))
\delta g_{\mu\nu}(\bm\xi)
\frac{\dot{z}^{\mu}(\lambda)
\dot{z}^{\nu}(\lambda)}{2l_{\bm\eta,\bm\eta'}}
\label{64}
\end{eqnarray}
where ${\bf z}(\lambda)={\bf z}_{\bm\eta,\bm\eta'}(\lambda)$ is
the geodesic which connects points $\bm\eta$ and $\bm\eta'$.
Substituting Eq.~(\ref{64}) into Eq.~(\ref{63}), and using the
definition of Eq.~(\ref{61}) we get the following representation
for the interaction part of the stress tensor
\begin{widetext}
\begin{equation}
\widetilde{W}^{\mu\nu}(\bm\xi,t) = -\frac{1}{2\sqrt{g}}
\int_{0}^{1}d\lambda\int d{\bm\eta}d{\bm\eta'}
\delta(\bm\xi - {\bf z}_{\bm\eta,\bm\eta'}(\lambda))
\frac{\dot{z}^{\mu}_{\bm\eta,\bm\eta'}(\lambda)
\dot{z}^{\nu}_{\bm\eta,\bm\eta'}(\lambda)}{l_{\bm\eta,\bm\eta'}}
\frac{\partial w(l_{\bm\eta,\bm\eta'})}{\partial l_{\bm\eta,\bm\eta'}}
\widetilde{\rho}_{2}(\bm\eta,\bm\eta')
\label{65}
\end{equation}
\end{widetext}
Let us evaluate the right hand side in Eq.~(\ref{65}) at Euclidean
metric $g_{\mu\nu}=\delta_{\mu\nu}$. In this case
$l_{\bm\eta,\bm\eta'}=|\bm\eta-\bm\eta'|$ while the geodesic
(parameterized by the natural parameter) is a straight line
$$
{\bf z}_{\bm\eta,\bm\eta'}(\lambda) = \bm\eta +
(\bm\eta' - \bm\eta)\lambda
$$
The above expressions for $l_{\bm\eta,\bm\eta'}$ and ${\bf
z}_{\bm\eta,\bm\eta'}(\lambda)$ should be substituted into
Eq.~(\ref{65}). Introducing a new variable $\bm\xi'=
\bm\eta'-\bm\eta$, and removing the delta-function by the
integration over $\bm\eta$ we obtain at the result that exactly
coincides with Eq.~(\ref{14}). Therefore the symmetric
representation of Eq.~(\ref{14}), which has been obtained in the
previous section by somewhat artificial manipulations, has a clear
geometric meaning. In particular, the internal parameter $\lambda$
in Eq.~(\ref{14}) is the natural parameter for a geodesic
connecting two interacting particles.

Equations (\ref{62}) and (\ref{65}) are the principal results of the
present subsection. They define explicit microscopic representations
for the stress tensors in a local noninertial reference frame.

The zero force condition of Eq.~(\ref{57}) with
$\widetilde{P}_{\mu\nu}$ of Eqs.~(\ref{62}) and (\ref{65}) is
equivalent to the requirement of zero current density,
Eq.~(\ref{47}). Hence we can use Eq.~(\ref{57}) as an alternative
``gauge'' condition to fix the velocity parameter $\widetilde{\bf
v}(\bm\xi,t)$, entering many-body equations of motion,
Eq.~(\ref{33}).

\section{Examples and applications}

\subsection{The harmonic potential theorem}
As a first simple example of application of our general formalism,
we consider a many-body dynamics in the presence of the following
external potential
\begin{equation}
U_{\text{ext}}({\bf x},t) = \frac{1}{2}m\omega_{\mu\nu}x^{\mu}x^{\nu}
+E_{\mu}(t)x^{\mu}
\label{66}
\end{equation}
where $\omega_{\mu\nu}$ is a constant tensor and $E_{\mu}(t)$ is a
time-dependent vector (without loss of generality we can put
$E_{\mu}(0)=0$). The initial value problem with the external
potential of Eq.~(\ref{66}) is exactly solvable, which is known as
the harmonic potential theorem (HPT) \cite{Dobson1994}. It is also
known that HPT is related to the covariance of the time-dependent
Sch\"odinger equation under the transformation to a global
accelerated reference frame \cite{Vignale1995a,Vignale1995b}.
Therefore our formulation of the many-body problem should be
perfectly suited to the demonstration of HPT.

Within the present Lagrangian formulation one needs to find a
selfconsistent solution to the many-body equation of motion,
Eq.~(\ref{33}), and to the force balance equation, Eq.~(\ref{57}).
Let us assume that velocity ${\bf v}({\bf x},t)$, which defines
(via Eq.~(\ref{18})) the motion of the reference frame, is a
function of $t$ only: ${\bf v}({\bf x},t) = {\bf V}(t)$. In this
case the trajectory of a fluid element takes a form
\begin{equation}
{\bf x}(\bm\xi,t)= \bm\xi + {\bf R}(t),
\label{67}
\end{equation}
where ${\bf R}(t)$ is a solution to the following Cauchy
problem
$$
\frac{\partial{\bf R}(t)}{\partial t} = {\bf V}(t),
\qquad {\bf R}(0) = 0
$$
Clearly, if our anzatz, ${\bf v}({\bf x},t) = {\bf V}(t)$, is a
selfconsistent solution, then ${\bf R}(t)$ should correspond to
the center-of-mass coordinate. Using Eq.~(\ref{67}) we get the
following results for the metric tensor $g_{\mu\nu}$, the velocity
$\widetilde{\bf
  v}(\bm\xi,t)$, and the effective potential, which enter
Eqs.~(\ref{33}) and (\ref{57})
\begin{eqnarray}
\label{68a}
 g_{\mu\nu} = \delta_{\mu\nu}, && \qquad
\widetilde{\bf v}(\bm\xi,t)={\bf V}(t)\\
\nonumber
U_{\text{ext}}({\bf x}(\bm\xi,t),t)
-&& m\frac{\widetilde{v}_{\nu}\widetilde{v}^{\nu}}{2} =
\frac{1}{2}m\omega_{\mu\nu}\xi^{\mu}\xi^{\nu} - L(t)\\
&& + \xi^{\mu}[m\omega_{\mu\nu}R^{\nu}(t) + E_{\mu}(t)]
\label{68}
\end{eqnarray}
Function $L(t)$ in Eq.~(\ref{68})
is the classical Lagrangian for a particle moving in the harmonic
potential of Eq.~(\ref{66})
$$
L(t) = \frac{m}{2}{\bf V}^{2}(t) -
\frac{1}{2}m\omega_{\mu\nu}R^{\mu}(t)R^{\nu}(t) -
E_{\mu}(t)R^{\mu}(t)
$$
The equation of motion, Eq.~(\ref{33}), and the force balance
equation, Eq.~(\ref{57}), simplify respectively as follows
\begin{widetext}
\begin{eqnarray}
&&i\frac{\partial\widetilde{\psi'}}{\partial t} =
\left(-\frac{\nabla_{\bm\xi}^{2}}{2m}
+ \frac{1}{2}m\omega_{\mu\nu}\xi^{\mu}\xi^{\nu}\right)
\widetilde{\psi'} + \int d{\bm\xi'}w(|{\bm\xi-\bm\xi'}|)
\widehat{\widetilde{n}}({\bm\xi'})\widetilde{\psi'}({\bm\xi}) +
\xi^{\mu}\left[m\frac{\partial V_{\mu}(t)}{\partial t}
+ m\omega_{\mu\nu}R^{\nu}(t) + E_{\mu}(t)\right]
\widetilde{\psi'}
\label{69} \\
&& m\frac{\partial V_{\mu}(t)}{\partial t}
+ m\omega_{\mu\nu}R^{\nu}(t) + E_{\mu}(t) + \left[
\frac{1}{n_{0}(\bm\xi)}\frac{\partial}{\partial\xi^{\nu}}
\widetilde{P}_{\mu\nu}(\bm\xi,t) + m\omega_{\mu\nu}\xi^{\nu}
\right] = 0
\label{70}
\end{eqnarray}
\end{widetext}
where we also performed a gauge transformation,
$\widetilde{\psi'}(\bm\xi,t)=\widetilde{\psi}(\bm\xi,t)
\exp[-im{\bf V}\bm\xi + i\int_{0}^{t}Ldt']$, which corresponds to the
transformation from the canonical to the kinematic momentum.

Let initially the system is prepared in a stationary state
(or in arbitrary mixture of stationary states). This means  that at $t=0$
the stationary force balance equation is fulfilled
\begin{equation}
\frac{1}{n_{0}(\bm\xi)}\frac{\partial}{\partial\xi^{\nu}}
\widetilde{P}_{\mu\nu}(\bm\xi,0) + m\omega_{\mu\nu}\xi^{\nu}=0.
\label{71}
\end{equation}
If at all $t>0$ the center-of-mass coordinate ${\bf R}(t)$ satisfies
the classical equation of motion
\begin{equation}
m\frac{\partial^{2} R_{\mu}(t)}{\partial t^{2}}
+ m\omega_{\mu\nu}R^{\nu}(t) + E_{\mu}(t) = 0,
\label{72}
\end{equation}
then both the equation of motion, Eq.~(\ref{69}), and the force
balance equation, Eq.~(\ref{70}), preserve their initial
(stationary) form. Therefore the many-body system in the co-moving
frame remains in the initial stationary state, in particular,
$\widetilde{P}_{\mu\nu}(\bm\xi,t)=\widetilde{P}_{\mu\nu}(\bm\xi,0)$.
This statement is the essence of HPT \cite{Dobson1994}. Within the
present formulation of many-body dynamics it appears quite
naturally. In fact, HPT is a built-in property of our
selfconsistent approach. This actually means that any approximate
treatment of a selfconsistent system of Eqs.~(\ref{33}),
(\ref{57}) should automatically satisfy HPT.

\subsection{Geometric formulation of generalized hydrodynamics:
  nonlinear elasticity of a collisionless Fermi gas}
  
The HPT type of motion provides an extremely simple example of many-body
dynamics without any deformation of local fluid elements
($g_{\mu\nu}^{\text{HPT}}=\delta_{\mu\nu}$). In this subsection we
apply our approach to a much more general situation with a nontrivial
dynamics of a fluid. Namely, we
consider a semiclassical dynamics of an interacting Fermi system in
the time-dependent Hartree approximation.
The problem reduces to the selfconsistent solution of a semiclassical
collisionless kinetic equation (see Eq.~(\ref{42}))
\begin{eqnarray}\nonumber
\frac{\partial \widetilde{f}'_{\bf K}}{\partial t} &+&
\frac{K^{\nu}}{m}\frac{\partial \widetilde{f}'_{\bf K}}{\partial
\xi^{\nu}} - \Big[m\frac{\partial \widetilde{v}_{\nu}}{\partial t}
+ K^{\mu}\widetilde{F}_{\mu\nu}
 - \frac{\partial g_{\alpha\beta}}{\partial \xi^{\nu}}
\frac{K^{\alpha}K^{\beta}}{2m} \\
&+& \frac{\partial}{\partial\xi^{\nu}}\left(U -
m\frac{\widetilde{v}_{\mu}\widetilde{v}^{\mu}}{2}\right)\Big]
\frac{\partial \widetilde{f}'_{\bf K}}{\partial K_{\nu}} = 0,
 \label{73}
\end{eqnarray}
and a force balance equation (see Eq.~(\ref{57}))
\begin{equation}
m\frac{\partial\widetilde{v}_{\mu}}{\partial t}
+ \frac{\partial}{\partial\xi^{\mu}}\left(U
- m\frac{\widetilde{v}_{\nu}\widetilde{v}^{\nu}}{2}\right)
+ \frac{\sqrt{g}}{n_{0}}\widetilde{P}^{\nu}_{\mu ; \nu} = 0.
\label{74}
\end{equation}
Here $U=U_{\text{ext}}(\bm\xi,t)+U_{\text{H}}(\bm\xi,t)$ is
a sum of the external potential and the Hartree potential,
$$
U_{\text{H}}(\bm\xi,t)=
\int w(l_{\bm\xi,\bm\xi'})n_{0}(\bm\xi')d\bm\xi'.
$$
Since the interaction effects are already included (on the mean field
level) in the selfconsistent potential, only kinetic part of the
stress tensor contributes to Eq.~(\ref{74}), i.e.
$\widetilde{P}_{\mu\nu} = m^{-1}\sum_{\bf
K}K_{\mu}K_{\nu}\widetilde{f}'_{\bf K}/\sqrt{g}$. The last expression
for $\widetilde{P}_{\mu\nu}$ is a plain semiclassical limit of the
general kinetic stress tensor, Eq.~(\ref{62}).

The problem of solving Eqs.~(\ref{73}), (\ref{74}) can be
reformulated as follows. Let us substitute the sum of the inertia and
the external forces from the balance equation, Eq.~(\ref{74}), into
the kinetic equation of Eq.~(\ref{73}). After the substitution
the potential $U$ in Eq.~(\ref{73}) cancels out and the kinetic
equation reduces to the following universal form
\begin{eqnarray}\nonumber
\frac{\partial \widetilde{f}'_{\bf K}}{\partial t} &+&
\frac{K^{\nu}}{m}\frac{\partial \widetilde{f}'_{\bf K}}{\partial
\xi^{\nu}}
 - \Big[ K^{\mu}\widetilde{F}_{\mu\nu}\\
 &-& \frac{\partial g_{\alpha\beta}}{\partial \xi^{\nu}}
\frac{K^{\alpha}K^{\beta}}{2m}
 - \frac{\sqrt{g}}{n_{0}}\widetilde{P}^{\mu}_{\nu ; \mu}\Big]
\frac{\partial \widetilde{f}'_{\bf K}}{\partial K_{\nu}} = 0
 \label{75}
\end{eqnarray}
where the stress tensor $\widetilde{P}_{\mu\nu}$ is defined as follows
\begin{equation}
\widetilde{P}_{\mu\nu}(\bm\xi,t) = \frac{1}{\sqrt{g}}
\sum_{\bf K}\frac{K_{\mu}K_{\nu}}{m}\widetilde{f}'_{\bf K}(\bm\xi,t)
\label{76}
\end{equation}
Equations (\ref{75}) and (\ref{76}) constitute a closed set, which
is structurally similar to the system of Vlasov equations with a
selfconsistent force. The skew-symmetric vorticity tensor,
$\widetilde{F}_{\mu\nu}(\bm\xi,t)$, and the symmetric deformation
tensor, $g_{\mu\nu}(\bm\xi,t)$, enter Eqs.~(\ref{75}) and
(\ref{76}) as external parameters, which govern the evolution of the
system. Hence these equations define a distribution function,
$\widetilde{f}'_{\bf K}(\bm\xi,t)$, as a unique functional of
$\widetilde{F}_{\mu\nu}$ and $g_{\mu\nu}$, provided the initial
condition, $\widetilde{f}'_{\bf
K}(\bm\xi,0)=\widetilde{f}^{(0)}_{\bf K}(\bm\xi)$, is given.
Equation (\ref{76}) determines the stress tensor as a universal
(i.~e. independent of external potential) functional of
$\widetilde{F}_{\mu\nu}$ and $g_{\mu\nu}$:
\begin{equation}
\widetilde{P}_{\mu\nu}=
\widetilde{P}_{\mu\nu}[\widetilde{F}_{\mu\nu},g_{\mu\nu}](\bm\xi,t).
\label{77}
\end{equation}
The vorticity and the deformation tensors contain nine independent
scalar functions (three from $\widetilde{F}_{\mu\nu}$ and six from
$g_{\mu\nu}$) which completely describe a deformed state of a
system \cite{PhysAc}. Hence Eq.~(\ref{77}) plays a role of a
generalized ``equation of state'' which relates the stress tensor
to the deformation. It is worth mentioning that the existence of
such an equation of state is a direct consequence of Runge-Gross
mapping theorem \cite{RunGro1984} in TDDFT.

Substituting the functional of Eq.~(\ref{77}) into Eq.~(\ref{74})
we obtain a hydrodynamic equation of motion which determines the
evolution of velocity for a given external potential. Therefore
the description of many-body dynamics consists of two separate
problems. The first one corresponds to the universal kinetic
problem of Eqs.~(\ref{75}), (\ref{76}). By solving these equations
we find the stress tensor functional, Eq.~(\ref{77}) (the
generalized equation of state). The second problem is to compute
the velocity and density distributions by solving the closed set
of hydrodynamics equations, Eqs.~(\ref{18}) and (\ref{74}).

The universal kinetic problem of Eqs.~(\ref{75}), (\ref{76}) can
be solved explicitly in the case of a fast long wavelength
dynamics, i.~e. if the deformation tensor is a fast function of
time, but slowly changes in space. More precisely, we assume that
the characteristic length scale, $L$, of the deformation
inhomogeneity is much large than $\tau u$, where $\tau$ is the
time scale of a dynamical process and $u$ is the characteristic
velocity of a particle. This situation is, for example, common in
Coulomb systems where the plasma frequency determines the
characteristic time scale of dynamics, while the corresponding
spatial variations of the density can be arbitrary slow. Let us
estimate different terms in Eq.~(\ref{75}) under the above
assumption. The first and the second terms in the right hand side
in Eq.~(\ref{75}) are of the order of $1/\tau$ and $u/L$
respectively. Both terms in the second line in Eq.~(\ref{75}) also
give a contribution $\sim u/L$, while the term, related to the
Coriolis force, is proportional to $\widetilde{F}\sim
\widetilde{v}_{T}/L$, where $\widetilde{v}_{T}$ is a rotational
(or transverse) component of the velocity. According to the force
balance equation of Eq.~(\ref{74}), for any physical velocity the
transverse part of the linear acceleration is compensated by the
transverse part of the vector
$\frac{\sqrt{g}}{n_{0}}\widetilde{P}^{\nu}_{\mu ; \nu}$ . Hence
$\widetilde{v}_{T}$ should be proportional to $\tau u^{2}/L$. This
means that the contribution of Coriolis force to the kinetic
equation is of the order of $\tau u^{2}/L^{2}$. Therefore, to the
leading order in the small parameter $\gamma =\tau u/L\ll 1$, only the
first term in Eq.~(\ref{75}) gives a nonvanishing contribution. Thus the
universal problem of Eqs.~(\ref{75}), (\ref{76}) reduces to the
following trivial equation
\begin{equation}
\frac{\partial}{\partial t}\widetilde{f}'_{\bf K}(\bm\xi,t) = 0.
\label{78}
\end{equation}
Equation (\ref{78}) shows that for a fast, small-gradient
evolution the distribution function in the Lagrangian frame preserves
its initial form: $\widetilde{f}'_{\bf
K}(\bm\xi,t)=\widetilde{f}^{(0)}_{\bf K}(\bm\xi)$. In this respect
the dynamics reminds the HPT type of motion. However the evolution
of the velocity is by far not trivial. Below we consider a system
which evolves from the equilibrium state. Substituting the
equilibrium distribution function into Eq.~(\ref{76}) we get the
stress tensor functional
\begin{equation}
\widetilde{P}_{\mu\nu}(\bm\xi,t) =
\frac{\delta_{\mu\nu}}{\sqrt{g(\bm\xi,t)}}P_{0}(\bm\xi),
\label{79}
\end{equation}
which is proportional to the initial equilibrium pressure,
$P_{0}(\bm\xi)$. The last step is to substitute the nonadiabatic
``equation of state'', Eq.~(\ref{79}), into the force balance
equation of Eq.~(\ref{74}). This results in the following
``hydrodynamic'' equation of motion
\begin{eqnarray}\nonumber
mn_{0}\frac{\partial\widetilde{v}_{\mu}}{\partial t}
&+& n_{0}\frac{\partial}{\partial\xi^{\mu}}\left(U
- m\frac{\widetilde{v}_{\nu}\widetilde{v}^{\nu}}{2}\right)\\
&+& \frac{\partial g^{\mu\nu}P_{0}}{\partial\xi^{\nu}}
+ \frac{1}{2}P_{0}\frac{\partial g^{\alpha\alpha}}{\partial\xi^{\mu}}  = 0,
\label{80}
\end{eqnarray}
where we used the definition of the covariant divergence,
Eq.~(\ref{56}), to compute the stress force,
$\sqrt{g}\widetilde{P}^{\nu}_{\mu ;\nu}$, in Eq.~(\ref{74}). We
would like to outline that $n_{0}(\bm\xi)$ and $P_{0}(\bm\xi)$ in
Eq.~(\ref{80}) are the time independent initial density and
pressure respectively. Equations (\ref{80}) and (\ref{18})
constitute a closed set of continuum mechanics equations which
describe a long wavelength dynamics of a Fermi gas in the
time-dependent Hartree approximation. Since the stress force in
Eq.~(\ref{80}) depends only on the deformation tensor,
$g_{\mu\nu}$, it is natural to interpret Eqs.~(\ref{80}) and
(\ref{18}) as a nonlinear elasticity theory of a Fermi gas. In the
case of small deformations this theory reduces to the standard
linear elasticity theory with a nonzero shear modulus. Indeed, in
the linear regime Eq.~(\ref{18}) takes the form
\begin{equation}
\frac{\partial {\bf u}({\bm\xi,t})}{\partial t} = {\bf  v}({\bm\xi,t}),
\label{81}
\end{equation}
where ${\bf u} = {\bf x}-{\bm\xi}$ is the displacement vector. The
deformation tensor reduces to the common linearized expression
\begin{equation}
g^{\mu\nu} = \delta_{\mu\nu} -
\frac{\partial u_{\mu}}{\partial\xi^{\nu}}
-\frac{\partial u_{\nu}}{\partial\xi^{\mu}}
\label{82}
\end{equation}
Assuming for simplicity that the unperturbed state is homogeneous, and
substituting Eqs.~(\ref{81}) and (\ref{82}) into Eq.~(\ref{80}), we get
the following equation of motion for the displacement vector
\begin{equation}
mn_{0}\frac{\partial^{2}u_{\mu}}{\partial t^{2}} -
\frac{\partial \sigma_{\mu\nu}}{\partial \xi^{\nu}} +
n_{0}\frac{\partial\delta U}{\partial \xi^{\mu}} = 0.
\label{83}
\end{equation}
The linearized stress tensor $\sigma_{\mu\nu}$ takes the standard
elastic form
\begin{equation}
\sigma_{\mu\nu} =
\delta_{\mu\nu}K\frac{\partial u_{\alpha}}{\partial\xi^{\alpha}} +
\mu\left(\frac{\partial u_{\mu}}{\partial\xi^{\nu}}
-\frac{\partial u_{\nu}}{\partial\xi^{\mu}} -
\delta_{\mu\nu}\frac{2}{3}
\frac{\partial u_{\alpha}}{\partial\xi^{\alpha}}\right),
\label{84}
\end{equation}
where  $K=\frac{5}{3}P_{0}$  and $\mu = P_{0}$ are the bulk
modulus and the shear modulus of a Fermi gas respectively
\cite{ConVig1999,TokPPRB1999,TokPPRB2000}.

Full nonlinear set of Eqs.~(\ref{80}), (\ref{18}) is equivalent to
the generalized collisionless hydrodynamics derived in
Refs.~\onlinecite{TokPPRB1999,TokPPRB2000} (see also 
Ref.~\onlinecite{AtwAshc2002}). In fact, Eqs.~(\ref{80}), (\ref{18}) 
and the generalized hydrodynamics of
Refs.~\onlinecite{TokPPRB1999,TokPPRB2000}  correspond to the same
theory in the Lagrangian and Eulerian formulations respectively. An
advantage of the present Lagrangian formulation is the
explicit form of the stress tensor $\widetilde{P}_{\mu\nu}$,
Eq.~(\ref{79}). The Lagrangian point of view also gives a very
clear microscopic picture of a fast collisionless dynamics. This is
a kind of evolution of a many-body system with almost
time-independent distribution of particles inside every moving and
deforming fluid element. Using the nonadiabatic equation of state
in the Lagrangian frame, Eq.~(\ref{79}), we can easily recover the
corresponding expression for the stress tensor $P_{\mu\nu}({\bf
x},t)$ in the laboratory frame
\begin{eqnarray}\nonumber
P_{\mu\nu}({\bf x},t) &=&
\frac{\partial \xi^{\alpha}}{\partial x^{\mu}}
\frac{\partial \xi^{\beta}}{\partial x^{\nu}}
\widetilde{P}_{\alpha\beta}(\xi({\bf x},t),t) \\
&=& \bar{g}_{\mu\nu}({\bf x},t)\sqrt{\bar{g}({\bf x},t)}
P_{0}(\xi({\bf x},t))
\label{85}
\end{eqnarray}
Here $\bar{g}_{\mu\nu}({\bf x},t)$ is Cauchy's deformation tensor
\cite{PhysAc}
\begin{equation}
\bar{g}_{\mu\nu}({\bf x},t) =
\frac{\partial \xi^{\alpha}}{\partial x^{\mu}}
\frac{\partial \xi^{\alpha}}{\partial x^{\nu}} , \qquad
\bar{g} = 1/g
\label{86}
\end{equation}
One can check that $P_{\mu\nu}({\bf x},t)$ of Eq.~(\ref{85}) is a
solution to the equation for the stress tensor derived in
Refs.~\onlinecite{TokPPRB1999,TokPPRB2000}. The stress tensor
$P_{\mu\nu}({\bf x},t)$, Eq.~(\ref{85}), enters the force balance
equation in the laboratory frame, Eq.~(\ref{16}). This is a
hydrodynamic equation of motion in Eulerian description. It is quite
natural that $P_{\mu\nu}({\bf x},t)$ depends on $\bar{g}_{\mu\nu}({\bf
x},t)$ since Cauchy's tensor is a common characteristics of
deformations in Eulerian picture.

In contrast to the stress tensor in the Lagrangian frame,
Eq.~(\ref{79}), the stress tensor of Eq.~(\ref{85}) is a highly
nonlocal function. It is proportional to the pressure $P_{0}$ at the
initial position of a fluid element which is currently at ${\bf x}$ 
(${\bf x}$
is an independent variable in Eq.~(\ref{16})). The locality of the
stress force in Eq.~(\ref{80}) is a key property of the present
Lagrangian formulation of a nonadiabatic continuum mechanics of a
Fermi gas. This formulations should be much more convenient for
applications to particular nonlinear problems.

\section{Conclusion}
We applied the idea of the Lagrangian description in continuum
mechanics to the theory of nonequilibrium quantum many-body
systems. Reformulation of the microscopic many-body theory in
terms of Lagrangian coordinates corresponds to the transformation
to the local noninertial reference frame moving with the flow (the
co-moving Lagrangian frame). This transformation allows to
separate the convective motion of particles, which is a direct
generalization of the common separation of the center-of-mass
motion in homogeneous systems. The motion of particles in the
Lagrangian frame is influenced by the external forces and by
generalized inertia forces. We have shown that the inertia forces
can be described in purely geometric terms of Green's deformation
tensor $g_{\mu\nu}$ and the skew-symmetric vorticity tensor
$\widetilde{F}_{\mu\nu}$. Tensors $g_{\mu\nu}$ and
$\widetilde{F}_{\mu\nu}$ enter equations of motion as an effective
metric tensor and an effective magnetic field respectively. Our
results demonstrate a close relation of the many-body dynamics
in Lagrangian frame to the quantum dynamics on curved manifolds.

We also derived local conservation laws for the number of
particles and for momentum in the Lagrangian frame, and presented closed
microscopic expressions for the stress tensor and for the
corresponding stress force. The local momentum conservation law in
the Lagrangian frame reduces to a zero force condition. The inertia
forces exactly compensate the external force and the stress force
in every point of the Lagrangian $\bm\xi$-space. The net force, exerted
on every fluid element, is exactly zero, which results in zero
current density and a time-independent density distribution. This
property is the main advantage of the Lagrangian description. It
suggests one of the most promising application of our formalism,
which is a new reformulation of TDDFT in a form similar to the
static theory. Indeed the main practical problem of TDDFT is an
inevitable strong nonlocality of exchange correlation potentials
\cite{Vignale1995a,Vignale1995b,TokPPRB2003}. The physical reason
for this is just the nonadiabatic motion of fluid elements. When
time is flowing, new and new fluid elements arrive at a given
point ${\bf x}$, and bring an information about surrounding space,
producing the above nonlocality. Using our reformulation of the
many-body theory as a basis for TDDFT one can completely remove
the very source on the nonlocality, which is of extreme practical
importance. In this paper we did not touch these questions since
it required an extended special consideration. A detailed
formulation of TDDFT in the Lagrangian frame will be presented in the
next paper of this series \cite{LagrFrameII}.

In this paper we also considered two illustrative examples of
application. The most interesting of them is the description a
nonlinear semiclassical dynamics of a collisionless Fermi gas. We
have shown that the full problem can be separated into two
independent parts. The first one is the solution of a universal
kinetic problem, which defines the stress tensor as a universal
functional of $g_{\mu\nu}$ and $F_{\mu\nu}$. This stress tensor is
used as an input for the second ``hydrodynamic'' part of the problem,
determining the dynamics of the velocity vector. This separation of
the initial many-body problem can be viewed as a particular
realization of TDDFT in the hydrodynamic formulation 
\cite{RunGro1984,TokPPRB2003}. In the case of a fast
long wavelength dynamics (similar to that for plasma
oscillations), the universal kinetic problem can be solved
explicitly. The solution is extremely simple -- the Wigner
function in the Lagrangian frame is time-independent. The
corresponding ``hydrodynamic'' problem also can be formulated in
the explicit form. It reduces to a closed nonlinear
elasticity theory of a Fermi gas. This elasticity theory is, in
fact, the Lagrangian formulation of the generalized hydrodynamics
derived in Refs.~\onlinecite{TokPPRB1999,TokPPRB2000}. The
generalized hydrodynamics proved to be useful in the description
of a small-amplitude collective dynamics of an electron gas. It gives the
correct (consistent with the kinetic treatment) dispersion of
plasma waves in  homogeneous systems
\cite{TokPPRB1999,TokPPRB2000} and recovers the exact dispersion
of the edge modes in a confined geometry \cite{TokPPRB2002}. The
results for the standing plasma waves in a parabolically trapped
electron gas are also quite reasonable \cite{DobLe2000,DobLe2002}.
The Lagrangian ``elasticity theory'' of a Fermi gas, derived in
Sec.~IVB, is structurally much more simple than the Eulerian
formulation of Refs.~\onlinecite{TokPPRB1999,TokPPRB2000}. Therefore, we
believe that it should provide a good basis for the description of
nonlinear dynamical effects in inhomogeneous many-electron systems.

This work was supported by the Deutsche Forschungsgemeinschaft under
Grant No. PA 516/2-3.

\appendix
\section{The divergence representation of the interaction stress force}

By definition the infraction stress force $\widehat{\bf F}^{\text{int}}$ is the commutator of the current operator $\widehat{\bf j}$ and the interaction Hamiltonian
$\widehat{W}$, 
\begin{equation}
\widehat{\bf F}^{\text{int}}({\bf x}) = 
im[\widehat{\bf j}({\bf x}),\widehat{W}]_{-}.
\label{A1a}     
\end{equation}
In the coordinate representation operators $\widehat{\bf j}$ and $\widehat{W}$ are defined as follows
\begin{eqnarray}
\widehat{\bf j}({\bf x}) &=& \frac{1}{2m}
\sum_{i}[\widehat{\bf p}_{i},\delta({\bf x} - {\bf x}_i)]_{+},
\label{A1b}\\
 \widehat{W} &=& \frac{1}{2}\sum_{i,j}w({\bf x}_i,{\bf x}_j),
\label{A1} 
\end{eqnarray}
where $i$ and $j$ label particles.      
Calculating the commutator of Eq.~(\ref{A1a}) we get the force in the
following form 
\begin{eqnarray} \nonumber
\widehat{\bf F}^{\text{int}}({\bf x}) = 
\frac{1}{2}\sum_{i,j}&&\Big[
\delta({\bf x}-{\bf x}_i)
\frac{\partial w({\bf x}_i,{\bf x}_j)}{\partial {\bf x}_i} \\
&& + \delta({\bf x}-{\bf x}_j)
\frac{\partial w({\bf x}_i,{\bf x}_j)}{\partial {\bf x}_j} 
\Big].
\label{A2}
\end{eqnarray}
If the interaction potential satisfies the Newton's third law,
\begin{equation}
\frac{\partial w({\bf x}_i,{\bf x}_j)}{\partial {\bf x}_i} =
- \frac{\partial w({\bf x}_i,{\bf x}_j)}{\partial {\bf x}_j},
\label{A3}
\end{equation}
Eq.~(\ref{A2}) takes the form
\begin{equation}
\widehat{\bf F}^{\text{int}}({\bf x}) = \frac{1}{2}\sum_{i,j}\left[
 \delta({\bf x}-{\bf x}_i) -     \delta({\bf x}-{\bf x}_j)
\right]\frac{\partial w({\bf x}_i,{\bf x}_j)}{\partial {\bf x}_i}.
\label{A4}
\end{equation}
The difference of delta-functions in Eq.~(\ref{A4}) can be transformed
as follows 
\begin{eqnarray}\nonumber
 \delta({\bf x}-{\bf x}_i) - \delta({\bf x}-{\bf x}_j) = 
        \left[1 - e^{({\bf x}_i-{\bf x}_j)\frac{\partial}{\partial
        {\bf x}}}\right]  
        \delta({\bf x}-{\bf x}_i) &&\\ \nonumber
= -({\bf x}_i-{\bf x}_j)\frac{\partial}{\partial {\bf
        x}}\int\limits_{0}^{1} d\lambda 
e^{\lambda ({\bf x}_i-{\bf x}_j)\frac{\partial}{\partial {\bf x}}}
\delta({\bf x}-{\bf x}_i) &&\\
= - \frac{\partial}{\partial {\bf x}}({\bf x}_i-{\bf x}_j)
\int\limits_{0}^{1}d\lambda 
\delta({\bf x}-{\bf x}_i - \lambda ({\bf x}_j-{\bf x}_i)).&& \qquad
\label{A5}
\end{eqnarray}
Inserting Eq.~(\ref{A5}) into Eq.~(\ref{A4}) we get the final
representation for the interaction stress force  
\begin{equation}
        \widehat{F}^{\text{int}}_{\mu}({\bf x}) = \frac{\partial}{\partial
        x^{\nu}}\widehat{W}_{\mu\nu}({\bf x}), 
\label{A6}
\end{equation}
where $\widehat{W}_{\mu\nu}({\bf x})$ is the interaction stress
tensor operator 
\begin{eqnarray}\nonumber
\widehat{W}_{\mu\nu}({\bf x}) = - \frac{1}{2}\sum_{i,j}
\int\limits_{0}^{1}d\lambda 
&&\delta({\bf x}-{\bf x}_i - \lambda ({\bf x}_j-{\bf x}_i))\\
&& \times (x_{i}^{\nu} - x_{j}^{\nu})
        \frac{\partial w({\bf x}_i,{\bf x}_j)}{\partial x_{i}^{\mu}}. \quad
\label{A7}
\end{eqnarray}
The $\lambda$-integration in Eq.~(\ref{A7}) is along the line that connects
two interacting particles.

% \bibliography{journals,bose,kinetics,tddft,hydrodynamics,geometry,eqDFT,mypapers,books,LagrFrameI_note}

\begin{thebibliography}{33}
\bibitem[{not({\natexlab{a}})}]{note1}
\bibinfo{note}{In Appendix~A we show that the interaction force is
  representable in a divergence form of Eq.~(\ref{13}) if the interaction
  potential, $w({\bf x},{\bf x'})$, satisfies the condition $\partial_{\bf x}
  w({\bf x},{\bf x'}) = - \partial_{\bf x'} w({\bf x},{\bf x'})$ that is the
  formal expression for the Newton's third law. If $w({\bf x},{\bf
  x'})=w(|{\bf{x}}-{\bf x'}|)$, the corresponding stress tensor is explicitly
  symmetric.}
\bibitem[{not({\natexlab{b}})}]{note2}
\bibinfo{note}{In classical continuum mechanics a similar tensor is sometimes
  called the spin tensor \cite{PhysAc}. Since in the present quantum context
  this term is somewhat misleading, we prefer to use the term "vorticity" to
  name the tensor $\widetilde{F}_{\mu\nu}$}.
\end{thebibliography}

\end{document}